\documentclass[sigconf]{acmart}

\usepackage{booktabs} 
\acmPrice{15.00}

\setcitestyle{square}

\begin{document}

\title{Lightform: Procedural Effects for Projected AR}

\author{Brittany Factura}
\affiliation{\institution{Lightform, Inc.}}
\author{Laura LaPerche}
\affiliation{\institution{Lightform, Inc.}}
\author{Phil Reyneri}
\affiliation{\institution{Lightform, Inc.}}
\author{Brett Jones}
\affiliation{\institution{Lightform, Inc.}}
\author{Kevin Karsch}
\affiliation{\institution{Lightform, Inc.}}

\renewcommand{\shortauthors}{Factura, LaPerche, Reyneri, Jones, and Karsch}


\copyrightyear{2018}
\acmYear{2018}
\setcopyright{rightsretained}
\acmConference{SIGGRAPH '18 Studio}{August 12-16, 2018}{Vancouver, BC, Canada}\acmDOI{10.1145/3214822.3214823}
\acmISBN{978-1-4503-5819-4/18/08}


\begin{teaserfigure}
  \vspace{-2mm}
  \centering
  \includegraphics[width=40mm]{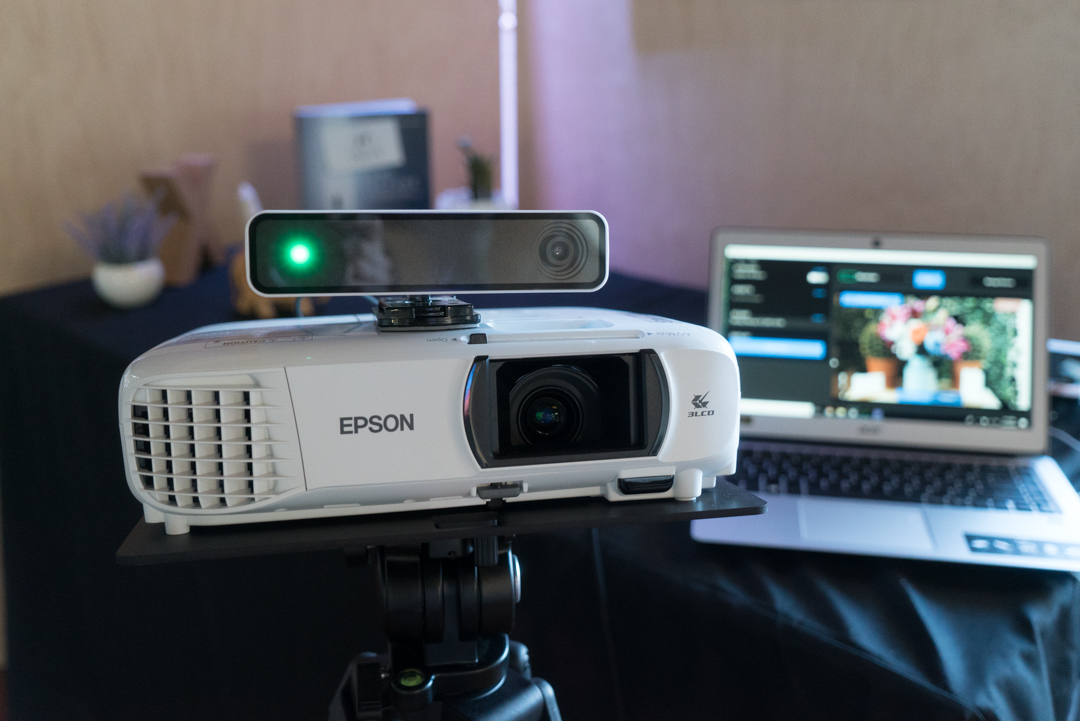}
  \includegraphics[width=40mm]{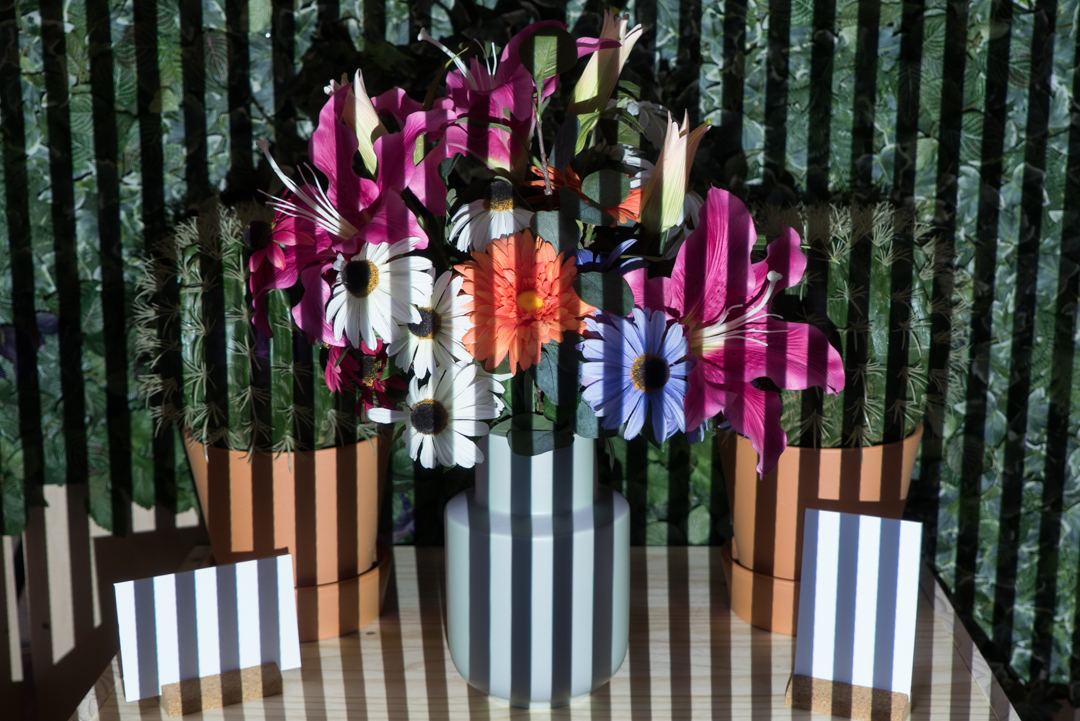}
  \includegraphics[width=47mm]{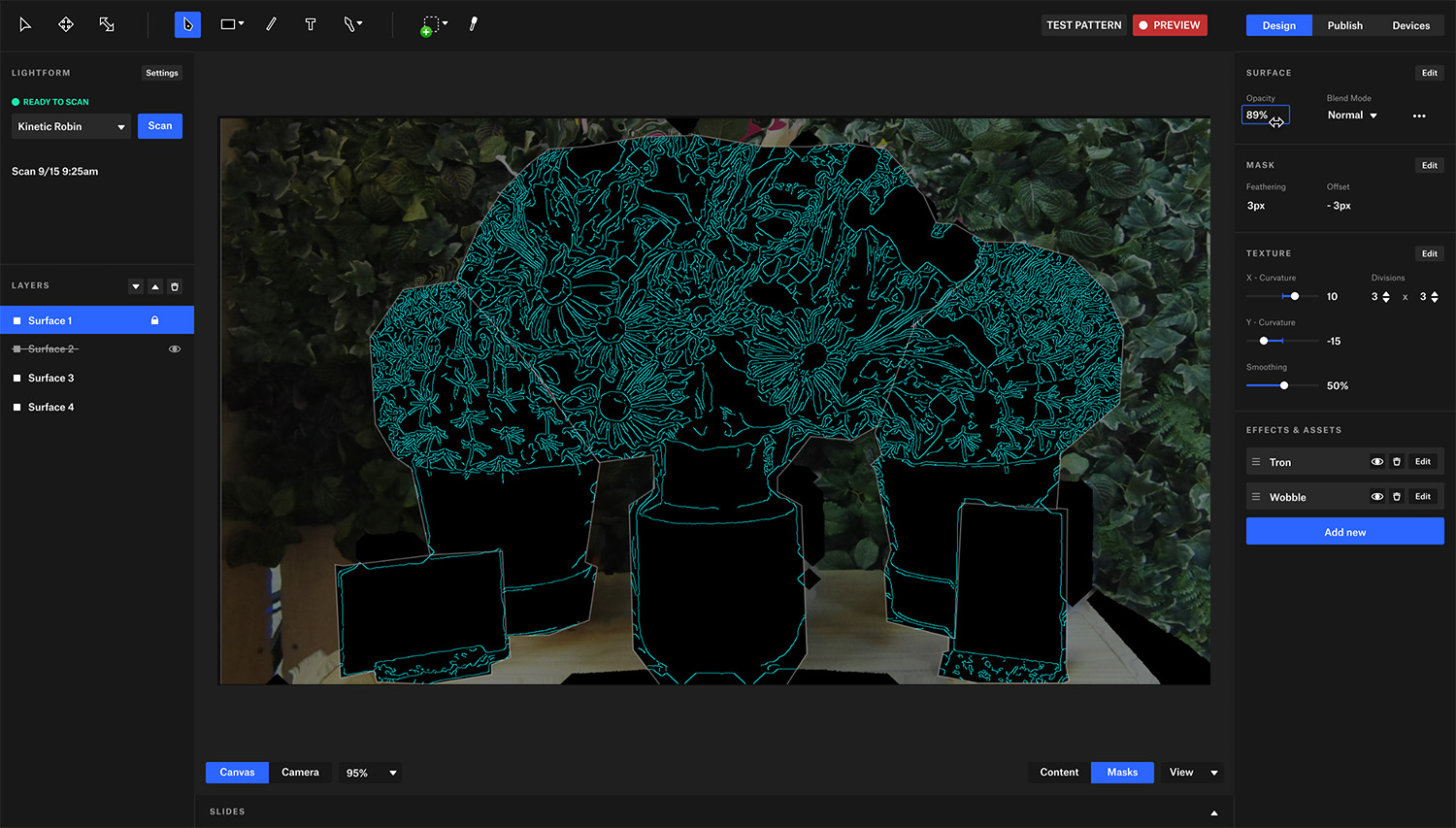}
  \includegraphics[width=40mm]{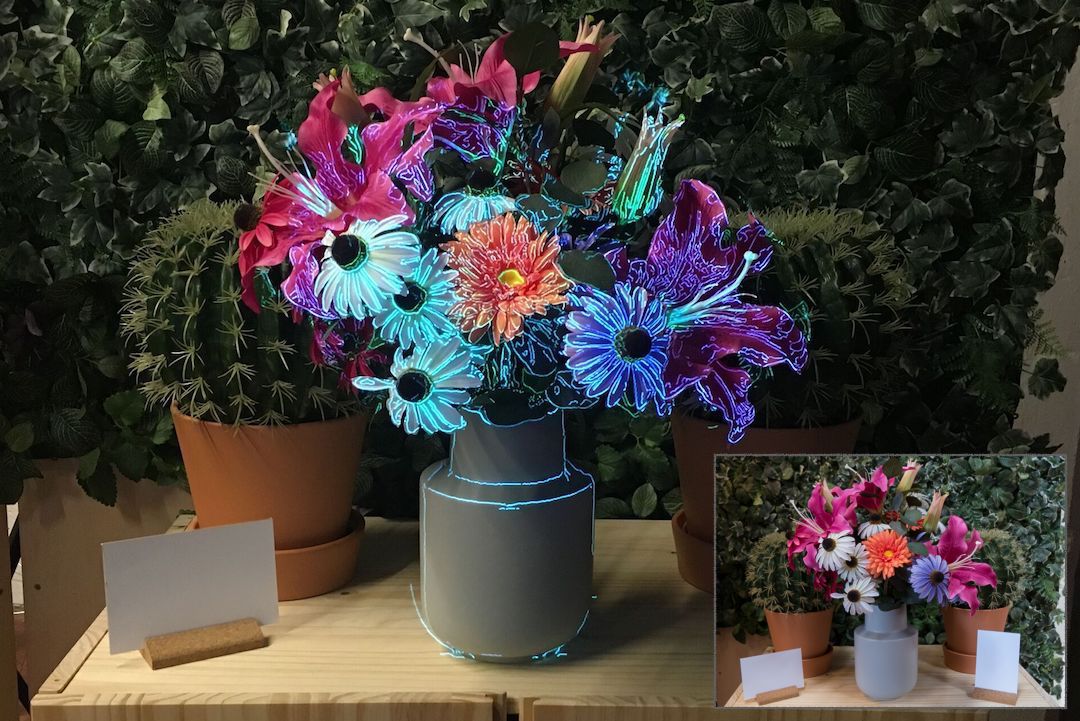}
  \vspace{-2.5mm}
  \caption{Lightform demonstration showing an automatic effect. From left to right: LF1 with laptop running Lightform Creator in the background; a scene during LF1 structured light scanning; screenshot of Lightform Creator during effect generation; the scene with the automatic "TRON" effect applied (original scene inset).}
  \vspace{1.0mm}
\end{teaserfigure}

\maketitle

\section{Introduction}
Projected augmented reality, also called projection mapping or video mapping, is a form of augmented reality that uses projected light to directly augment 3D surfaces, as opposed to using pass-through screens or headsets. The value of projected AR is its ability to add a layer of digital content directly onto physical objects or environments in a way that can be instantaneously viewed by multiple people, unencumbered by a screen or additional setup.

Because projected AR typically involves projecting onto non-flat, textured objects (especially those that are conventionally not used as projection surfaces), the digital content needs to be mapped and aligned to precisely fit the physical scene to ensure a compelling experience. Current projected AR techniques require extensive calibration at the time of installation, which is not conducive to iteration or change, whether intentional (the scene is reconfigured) or not (the projector is bumped or settles). The workflows are undefined and fragmented, thus making it confusing and difficult for many to approach projected AR. For example, a digital artist may have the software expertise to create AR content, but could not complete an installation without experience in mounting, blending, and realigning projector(s); the converse is true for many A/V installation teams/professionals. Projection mapping has therefore been limited to high-end event productions, concerts, and films, because it requires expensive, complex tools, and skilled teams (\$100K+ budgets).

Lightform provides a technology that makes projected AR approachable, practical, intelligent, and robust through integrated hardware and computer-vision software. Lightform brings together and unites a currently fragmented workflow into a single cohesive process that provides users with an approachable and robust method to create and control projected AR experiences.


\section{Lightform Design and Implementation}

Lightform is comprised of two solutions: a small camera/computer device called the "LF1" which physically attaches to a projector, and a desktop software application called "Lightform Creator" that allows a user to control the LF1 and also create projected AR content that can be uploaded and played through the LF1. Using computer vision and machine learning, we automate and simplify many of the pain points associated with projected AR, such as (re)alignment and content creation. We believe that these tools will help make projected AR faster, easier and cheaper than before.

\subsection{Lightform LF1}
The LF1 contains mobile processors as well as a 12 megapixel (4096x3072) RGB imaging sensor. It can be connected to a network over WiFi or Ethernet, and has an HDMI port for connecting to most any projector. Modern projectors come in a variety of throw ratios (field of view), and the LF1 supports different lens configurations to support many of these.

The LF1 acts as a media playback and hosting server, but most importantly, it is responsible for acquiring structured light scans of the scene, and computing procedural\footnote{Procedural, in this sense, implies that effects adapt to the contents (color and/or geometry) of a given scene {\it automatically}.} AR content (effects). The data acquired by the LF1 is also sent to Lightform Creator for users who wish to create more complex AR experiences.

\subsubsection{Structured Light Scanning}
Structured light scanning is one of the first and most important steps of the Lightform workflow, enabling both realignment and content creation. We have implemented a visible structured light algorithm (inspired by Yamazaki et al.~\cite{Yamazaki:2011}) on the LF1 for remotely capturing detailed scene information. A structured light scan operates by projecting patterns which are captured by a camera (in this case, the LF1). This provides a dense pixel correspondence from projector pixels to/from camera pixels. This mapping can be thought of in much the same way as stereo camera calibration and reconstruction~\cite{Hartley:2003}, and we apply similar techniques to extract 3D information (disparity and depth) from the scene. One important benefit that this provides is the ability to reconstruct a {\it projector image}, in other words, a remapping of camera pixels into the projector's domain to obtain an image of the scene as if taken with the projector's optical parameters and point-of-view. The projector image is extremely important for creating procedural effects, as it provides an image that can be understood and processed by vision techniques (e.g. segmentation, edge detection, etc) and effect-generating algorithms. A user can author content directly on top of this image, and it is automatically aligned with the real world, eliminating the need for traditional mapping workflows.


\subsubsection{Procedural Effects: Real World Image Filters}
Lightform uses the structured light scan data to create instant effects, which makes it easy to produce complex, dynamic motion content without motion graphics or multimedia expertise. Using the {\it projector image}, effects are created by applying various image processing and computer vision techniques that result in interesting and unique animations that adapt to the scene. For example, the "TRON" effect traces edges present in the scene. This effect first applies Canny edge detection to the projector image, then uses a shader to perform edge tracing on the detected edges. The result is an animation in which the edges of the scene glow and change over time. We show several of these effects in the accompanying video, and we think of these as AR filters (in the image processing sense) for the real world.

{\bf Shadertoy}. Besides the set of premade effects we have created, the LF1 supports Shadertoy formatted effects (GLSL shaders that utilize the same structure and syntax as those found on \url{http://www.shadertoy.com}). This enables custom effects that can be created by users and shared among the community, or simply downloaded from Shaderytoy and applied as a projected AR effect.

\subsection{Lightform Creator}
Lightform Creator is a cross-platform desktop software application. It can control the LF1 (trigger a scan, stream video/camera images, play/pause content, etc), but most importantly, it is a tool for easily creating projected AR content. This tool can be described as a hybrid between 2D multimedia editing software (e.g. Adobe's After Effects) and 3D modeling software (e.g. Autodesk's Maya), however our aim is to simplify the user interactions through the use of machine learning and computer vision. For the purposes of this paper, we will focus on Lightform Creator's procedural effects capabilities.

After capturing a scan and receiving the projector image from the LF1, Lightform Creator allows a user to quickly create masks (regions where content should be projected in a scene) with vision-assisted tools similar to Photoshop's Quick Select, Magic Wand and Magnetic Lasso tools. Once masks have been created, the user can choose from a list of procedural effects to be applied to the scene. For example, TRON traces the edges of a scene, another distorts a scene to attract attention, and another makes the scene appear as if it were a hand-drawn cartoon; many effects are inspired by the Illumiroom project~\cite{Jones:2013}. See the accompanying video for a demonstration.

\section{Applications of Lightform}
In the past, projected AR has been demonstrated primarily in the academic community and in high-budget concerts and shows. Lightform encourages and enables these as well as many new AR applications; different from existing AR technology, these do not require headsets or additional hardware, and can be consumed by multiple viewers at the same time.

We are focused on allowing projected AR experiences to be shared, friction-less, cost-effective, and hidden by presenting an out-of-home technology that can be scaled to many viewers and viewed naturally with the naked human eye. We believe in the importance of replacing printed signs or television screens to overlay digital information onto the real world using a technology that is invisible, seamless, and unencumbered by screens, thus providing an out-of-home digital display experience that is novel to those that experience it. Digital data, art, and "magic" can be seamlessly integrated into everyday spaces.

We are particularly interested in using Lightform to communicate and display information or attract attention towards a specific object or venue (digital signage), create stand-alone art pieces (ambiance), design a space (place-making and identity), or purely to provide entertainment and joy (art/performance). Specifically, a range of projects can be enabled with projected AR: custom digital content can augment and transform a mural; a cafe menu can be easily animated with dynamic and updatable items without affecting the aesthetic of a restaurant; compelling effects can animate onto a 3D sculpture; in general, spaces can be transformed completely, allowing viewers to engage with their surroundings in new and meaningful ways.


\section{SIGGRAPH Demonstration}
We will demonstrate an interactive tabletop where visitors will be able to customize their own projected AR scene using various physical objects and a drawing station. The LF1 will scan the setup the visitors have created, and users will then have the ability to apply effects onto their scene. We will also demonstrate a second LF1 setup running premade effects onto a variety of art pieces and sculptures within the booth.


\bibliographystyle{ACM-Reference-Format}
\bibliography{refs}
\end{document}